\documentclass[12pt]{iopart}
\usepackage{graphicx}
\begin{document}
\hfill  DAMTP-2014-80

\title[]{Minimal Massive 3D Gravity Unitarity Redux}

\author{Alex S. Arvanitakis and Paul K. Townsend}

\address{Department of Applied Mathematics and Theoretical Physics,\\ Centre for Mathematical Sciences, University of Cambridge,\\
Wilberforce Road, Cambridge, CB3 0WA, U.K.}
\ead{A.S.Arvanitakis@damtp.cam.ac.uk, P.K.Townsend@damtp.cam.ac.uk}
\vspace{10pt}
\begin{abstract}
A geometrical analysis of  the bulk and anti-de Sitter boundary unitarity conditions of 3D ``Minimal Massive Gravity'' (MMG)
(which evades the ``bulk/boundary clash'' of Topologically Massive Gravity) is used to extend and simplify previous results, showing that unitarity selects, up to equivalence, a connected region in parameter space. We also initiate the study of flat-space holography for MMG. Its relevant flat space limit is  a deformation of 3D conformal gravity; the deformation is both non-linear and non-conformal, implying a linearisation instability. 
\end{abstract}

%
%
%
%
%

\section{Introduction}

A recently proposed model of 3D massive gravity \cite{Bergshoeff:2014pca}, dubbed ``Minimal Massive Gravity'' (MMG),  has bulk properties that are identical to those of  ``Topologically Massive Gravity'' (TMG) (which propagates a single massive spin-$2$ mode \cite{Deser:1981wh})  but its boundary properties (for AdS asymptotics) are different.  Specifically, MMG evades the ``bulk/boundary clash'' of TMG; this is the impossibility (for TMG) of arranging for both central charges  of the asymptotic conformal symmetry algebra to be positive while also arranging for the bulk mode to have positive energy.

In this paper we present a greatly simplified, and geometrical,  analysis of the unitarity conditions of MMG.  Our results confirm those  of  \cite{Bergshoeff:2014pca} but we also consider a slightly larger class of models by leaving free the normalisation of the parameters of the MMG action, and we cut in half the relevant parameter  space by establishing  equivalence under a ``duality'' transformation in the full parameter space. The final result is that unitarity restricts the parameters to a connected region of parameter space, up to equivalence. 

It has been proposed \cite{Bagchi:2009pe,Bagchi:2012cy} that the relevant  asymptotic symmetry algebra for 3D gravity with flat-space asymptotics is the  2D ``Galilean Conformal Algebra'' (GCA), obtained by contraction of the 2D conformal algebra. Based on this proposal, flat space holography for TMG was initiated in  \cite{Bagchi:2012yk}, and it was argued that unitarity constraints on the central charges of the GCA algebra  could be satisfied only in a limit for which TMG degenerates to 3D conformal gravity, which has no local degrees of freedom. Here we show that the same analysis, applied to MMG, leads to a flat space 
MMG model that is not conformal and still has local degrees of freedom, although its linearised limit coincides with linearised 3D conformal gravity, implying a linearisation instability. 

\section{TMG preliminaries}

 The TMG action  can be written as the integral of a Lagrangian 3-form constructed from three Lorentz-vector one-forms: the dreibein $e$, the (dual) Lorentz connection $\omega$ and a Lagrange multipler field $h$ imposing a zero-torsion constraint \cite{Grumiller:2008pr,Carlip:2008qh}. Using a 3D vector algebra notation for Lorentz vectors we can write this Lagrangian 3-form as 
\begin{equation}
L_{TMG}[e,\omega,h] = -\sigma e\cdot R + \frac{1}{6}\Lambda_0\,  e\cdot e\times e +  h\cdot T  + \frac{1}{\mu} L_{LCS}(\omega)\, , 
\end{equation}
where $T$ and $R$ are the torsion and curvature 2-forms, respectively, and $L_{LCS}$ is the Lorentz-Chern-Simons (LCS) 3-form for $\omega$. This Lagrangian 3-form is parametrized by the mass parameter $\mu$ and the dimensionless constants $(\sigma,\Lambda_0/\mu^2)$. We refer the reader to \cite{Bergshoeff:2014pca} for details of conventions.  

The 1-form fields $(\omega,h)$ can be eliminated by their joint equations of motion, leading to an action for the 3-metric alone. For a particular choice of units for the 3D Newton constant $G_3$ (which has dimensions of inverse mass for unit speed of light),  the resulting action is
\begin{equation}
I_{TMG}[g] = \sigma I_{EH}[g;\Lambda] + \frac{1}{\mu} I_{LCS}[g]\, , 
\end{equation}
where $I_{EH}$ is the (3D) Einstein-Hilbert action, together with a cosmological constant term ($\sigma\Lambda= \Lambda_0$);  the LCS term is now the Chern-Simons action for the  Levi-Civita affine connection.  Variation of this action yields the TMG equation
\begin{equation}
\sigma \left(G_{\mu\nu} + \Lambda g_{\mu\nu} \right) + \frac{1}{\mu}\, C_{\mu\nu} =0\, ,
\end{equation}
where $G$ is the Einstein tensor and $C$ the Cotton tensor.  

The asymptotic symmetry algebra for an asymptotic AdS  vacuum with cosmological constant $\Lambda=-1/\ell^2$  is the 2D conformal algebra \cite{Brown:1986nw}. For TMG the  (left/right) central charges are
\begin{equation}\label{cpm}
c_\pm = \frac{3\ell}{2G_3} \left(\sigma \pm \frac{1}{\mu\ell}\right)\, .
\end{equation}
In the context of an AdS/CFT interpretation, this is the leading order result in a semi-classical approximation that is valid (re-instating factors of $\hbar$) when
$\ell/G_3 \gg \hbar$, and quantum consistency requires $8\mu G_3=1/N$ for some integer $N$ (see e.g. \cite{Bagchi:2012yk}). 

We may assume, without loss of generality, that 
\begin{equation}
\mu\ell > 0\, . 
\end{equation}
Given this,  we see from (\ref{cpm}) that $c_-$ will be negative and the (putative) dual CFT non-unitary, unless $\sigma>0$. However, the spin-2 bulk mode is a ``ghost'' (i.e. has negative energy) unless $\sigma<0$. This is the bulk/boundary unitarity clash.

 \section{MMG}

 The MMG action cannot be written in terms of the metric alone but it can be written in terms of the $1$-form Lorentz-vector fields $(e,\omega,h)$. The Lagrangian 3-form is \cite{Bergshoeff:2014pca}
 \begin{equation}\label{MMGact}
 L_{MMG}[e,\omega,h]= L_{TMG}[e,\omega,h]  + \frac{\alpha}{2} e\cdot h\times h\, , 
 \end{equation}
 where $\alpha$ is a new dimensionless constant; for reasons explained in  \cite{Bergshoeff:2014pca}, the parameters are restricted such that 
 \begin{equation}
 1+ \sigma\alpha \ne0\, . 
 \end{equation}
  The 1-form fields $(\omega,h)$ cannot be consistently eliminated from the action but the complete set of equations of motion 
 determine them uniquely in terms of the dreibein and its derivatives. When the resulting equation of motion is expressed in terms of the metric, one finds that 
  \begin{equation}\label{sourcefree}
\bar\sigma G_{\mu\nu} +  \bar\Lambda_0\,  g_{\mu\nu}  + \frac{1}{\mu} C_{\mu\nu} + \frac{\gamma}{\mu^2} J_{\mu\nu} =0\, , 
\end{equation}
where\footnote{We use here the version of these formulae given in \cite{Arvanitakis:2014yja}, which are valid for any value of $\sigma$.} 
\begin{equation}\label{bars}
\bar\sigma = \sigma(1+\sigma\alpha) + \frac{\alpha^2\Lambda_0}{2\mu^2(1+\sigma\alpha)^2} \, , \qquad \gamma= - \frac{\alpha}{(1+\sigma\alpha)^2}\, , 
\end{equation}
and 
\begin{equation}\label{barlamzero}
\bar\Lambda_0 = \Lambda_0\left(1+\sigma\alpha - \frac{\alpha^3\Lambda_0}{4\mu^2(1+\sigma\alpha)^2}\right)\, . 
\end{equation}
The $J$-tensor is 
\begin{equation}\label{Jeq}
J_{\mu\nu} =  -S_\mu{}^\rho S_{\rho\nu} + S S_{\mu\nu} + \frac{1}{2}g_{\mu\nu}\left(S^{\rho\sigma}S_{\rho\sigma}-S^2\right)\, , 
\end{equation}
where  $S_{\mu\nu}= R_{\mu\nu} - (1/4)g_{\mu\nu} R$ (the 3D Schouten tensor) and the scalar $S$ is its trace. Notice that the $J$-tensor involves only second derivatives of the metric, so the characteristics of the MMG equation will be those of the TMG equation, found by setting $\gamma=0$. We defer to \cite{Bergshoeff:2014pca} for an explanation of how the MMG equation manages to be consistent with Bianchi identities despite the fact that $D_\mu J^{\mu\nu}\not\equiv 0$.

\subsection{Scaling parameters}

The MMG action is proportional to the integral of the Lagrangian 3-form (\ref{MMGact}) with a proportionality constant that is itself proportional to $1/G_3$. A rescaling of $G_3$ can be compensated by a rescaling of the MMG parameters.  Specifically, if $G_3\to \lambda G_3$, for {\it positive non-zero} constant $\lambda$, then the action is unchanged if
\begin{equation}
\mu\to \lambda^{-1}\mu\, ,  \qquad \alpha \to \lambda^{-1} \alpha\, , \qquad \sigma\to \lambda\sigma\, , \qquad \Lambda_0\to \lambda \Lambda_0\, , 
\end{equation}
and 
\begin{equation}
h \to \lambda h  \, , \qquad (e,\omega)  \to (e,\omega)\, . 
\end{equation}
Given that $\sigma\ne0$, this scaling allows us to choose a normalisation for the parameters such that $\sigma^2=1$, and this choice was made in \cite{Bergshoeff:2014pca}. For TMG it makes sense to 
make this choice since setting $\sigma=0$ leads to inconsistent field equations unless one also sets $\Lambda_0=0$ and then one is left with 3D conformal gravity, which has no local degrees of freedom. 
For MMG it is perfectly consistent to set $\sigma=0$, and allowing for this case leads to significant simplifications. For this reason, we impose no  restriction on the parameter 
$\sigma$. 

We could use the scaling symmetry to set $\alpha^2=1$. This is especially attractive since unitarity of MMG (with AdS asymptotics) requires $\alpha<0$ \cite{Bergshoeff:2014pca} so, anticipating this result, we could restrict to $\alpha=-1$. However, imposing a normalisation condition on $\alpha$ would complicate any discussion of the TMG limit.  For this reason, we shall proceed without making any choice of normalisation for the MMG parameters. 

\subsection{A duality in parameter space}\label{subsec:duality}

Let us define a new set of fields $(\hat e,\hat\omega,\hat h)$ for the action (\ref{MMGact}) by setting
\begin{equation}
e= \hat e \, , \qquad \omega= \hat \omega + 2 \mu \frac{(1 + \sigma \alpha)}{\alpha } \hat e\, , \qquad 
h = \hat h - 2 \mu \frac{(1 + \sigma \alpha)}{\alpha^2} \hat e \, . 
\end{equation}
This is an invertible field redefinition and therefore has no physical effect. However, the action in terms of the new fields has exactly the same form as it had for the old fields but in terms of the new parameters 
\begin{equation}
\hat\sigma = -(2+\sigma\alpha)/\alpha \, , \qquad 
\hat\Lambda_0 =  \Lambda_0 - \frac{4(1+\sigma\alpha)^3}{\alpha^3} \mu^2 \, , \qquad 
\hat\alpha = \alpha\, . 
\end{equation}
In other words,  the map
\begin{equation}
H: \ (\sigma,\Lambda_0,\alpha) \to (\hat\sigma,\hat\Lambda_0,\hat\alpha)
\end{equation}
takes one MMG model into an equivalent MMG model. In particular, the MMG field equation (\ref{sourcefree}) is unchanged because the parameters $(\bar\sigma,\gamma,\bar\Lambda_0)$ are $H$-inert. 

Notice that the $H$-map is an involution ($H^2=1$) so it relates {\it pairs} of equivalent models; it is a ``duality'' in parameter space. 
Notice also that 
\begin{equation}
H:\ (1+\sigma\alpha)\ \to\  - (1+\sigma\alpha)\, . 
\end{equation}
Recall that $(1+\sigma\alpha)\ne0$ by definition, so $H$ has no fixed points.  It follows that we may choose $(1+\sigma\alpha)$ to have either sign, without loss of generality. Every MMG model for which $(1+\sigma\alpha)<0$ is equivalent to one with  $(1+\sigma\alpha)>0$. 

This result raises a puzzle. When $(1+\sigma\alpha)>0$ we can take the $\alpha\to0$ limit to recover TMG from MMG. Given equivalence under the action of H, there should also be a limit to TMG when $(1+\sigma\alpha)<0$,  but this cannot be a simple $\alpha\to0$ limit because this would change the sign of $(1+\sigma\alpha)$. The resolution of this puzzle is that there is indeed another, less obvious, TMG limit when $(1+\sigma\alpha)<0$, which may be deduced by 
following the consequences of the $H$-map.  One takes
\begin{equation}
\alpha\to 0\, , \qquad \sigma\to \infty\, , \qquad \Lambda_0\to \infty\, , 
\end{equation}
in such a way that for,  non-zero $\alpha$, 
\begin{equation}
\sigma= -\frac{1}{\alpha} \left(2+\hat\sigma\alpha\right)\, , \qquad \Lambda_0/\mu^2 = -\frac{4(1+\hat\sigma\alpha)^3}{\alpha^3} + \hat\Lambda_0/\mu^2\, , 
\end{equation}
where $\hat\sigma$ and $\hat\Lambda_0$ are finite in the $\alpha\to0$ limit. This limit now yields a TMG model with 
\begin{equation}
\bar\sigma = \hat\sigma\, , \qquad \bar\Lambda_0 = \hat\Lambda_0\, .
\end{equation}

\section{MMG unitarity redux}

Maximally symmetric solutions of the MMG equation (\ref{sourcefree}) are characterized by the value of the cosmological constant $\Lambda$, which can be defined (in the context of a vacuum solution) via the equation $G_{\mu\nu}=-\Lambda g_{\mu\nu}$. Substitution into the MMG equation yields the relation 
\begin{equation}\label{quad}
\gamma \Lambda^2  -4\mu^2\bar\sigma \Lambda + 4\mu^2\bar\Lambda_0 =0\, . 
\end{equation}
In the case of an AdS vacuum we have a unitarity constraint arising from the requirement of positive central charges for the asymptotic 2D conformal algebra, in addition to the bulk unitarity constraints. The unitarity conditions found  for MMG in \cite{Bergshoeff:2014pca} are
\begin{itemize}

\item No-tachyon condition:
\begin{equation}\label{notach}
\sigma^2 + \frac{\left(\Lambda+ \alpha\Lambda_0\right)}{\mu^2(1+\sigma\alpha)^2} >0\, . 
\end{equation}

\item  No-ghost condition. When combined with the no-tachyon condition this  becomes the inequality 
\begin{equation}\label{ineq2}
\sigma(1+\sigma\alpha) + \frac{\alpha\left(\Lambda+\alpha\Lambda_0\right)}{2\mu^2(1+\sigma\alpha)^2} <0\, . 
\end{equation}

\item Positivity of both central charges:
\begin{equation}
\sigma - \frac{1}{\mu\ell} - \frac{\alpha\left(\Lambda + \alpha\Lambda_0\right)}{2\mu^2(1+\sigma\alpha)^2} >0\, . 
\end{equation}
This inequality is valid on the assumption (which we make without loss of generality) that $\mu\ell>0$ (recall that $\Lambda = -1/\ell^2$) and saturation of it defines, in analogy to TMG \cite{Li:2008dq}, the ``chiral'' limit.  Properties of chiral MMG have been investigated recently in \cite{Tekin:2014jna,Alishahiha:2014dma}.

\end{itemize}
For TMG it is obvious that the chiral limit implies saturation of the no-tachyon condition. This is no longer obvious for MMG but it is true nevertheless, as will emerge from the analysis to follow. 

\subsection{Geometrical formulation of unitarity conditions}

We now define new dimensionless variables $(x,y)$ by
\begin{equation}\label{xandy}
x =  - \frac{\left(\Lambda+\alpha\Lambda_0\right)}{\mu^2(1+\sigma\alpha)^2}\, , \qquad y= \frac{2}{\mu\ell} \, .
\end{equation}
Observe that the inequality $y>0$ follows immediately from our assumption that $\mu\ell >0$.  It also follows that 
\begin{equation}
y^2- \alpha^2 x^2 -4(1+\sigma\alpha) x \equiv - \frac{\gamma}{\mu^4}\left(\gamma\Lambda^2 -4\mu^2\bar\sigma \Lambda + 4\mu^2\bar\Lambda_0\right) =0
\end{equation}
where the final equality follows from (\ref{quad}).  The new notation also simplifies the result of \cite{Bergshoeff:2014pca} for the central charges of the asymptotic 2D conformal algebra, which can now be written as
\begin{equation}\label{cpm1}
c_\pm = \frac{3\ell}{4G_3} \left(2\sigma + \alpha x \pm y \right)\, . 
\end{equation}

The unitarity conditions can now be written as linear inequalities in the $(x,y)$-plane:
\begin{itemize}

\item No-tachyon: $x<\sigma^2$.

\item No-ghost:  $-\alpha x + 2\sigma(1+\sigma\alpha) <0$. 

\item Positive central charges: $y<2\sigma + \alpha x$. 

\end{itemize}

To summarise, the ``allowed''  region in the $(x,y)$-plane is defined by the linear inequalities
\begin{equation}\label{ineq4}
x<\sigma^2\, , \qquad -\alpha x + 2\sigma(1+\sigma\alpha) <0\, , \qquad  0<y<2\sigma +\alpha x\, . 
\end{equation}
Within this allowed region, each point on the ``MMG hyperbola'', 
\begin{equation}\label{MMGhyper}
y^2 = \alpha^2 x^2 + 4(1+\sigma\alpha) x\, , 
\end{equation}
corresponds to an MMG model with AdS$_3$ vacuum satisfying all (bulk and boundary) unitarity conditions. 

Observe that the boundary lines $x=\sigma^2$   and $y=2\sigma+ \alpha x$ meet {\it on the hyperbola} at the ``chiral'' point with coordinates
\begin{equation}
x_{\rm ch}= \sigma^2\, , \qquad y_{\rm ch}= \sigma(2+\sigma\alpha)\, .
\end{equation}
This confirms our earlier claim that the no-tachyon condition for MMG is saturated at the critical point.  The other intersection of the two lines $x=\sigma^2$ and
 $y=2\sigma+ \alpha x$ is also on the hyperbola, but this corresponds to $c_+=0$, and hence to  $c_-<0$. So boundary unitarity is violated at this other chiral point, although it is also true (as we shall see shortly) that some unitarity condition is violated in any neighbourhood of either  chiral point.

\subsection{TMG redux}

We will first use this new geometrical framework to recover standard results for TMG. We can do this by setting $\alpha=0$, in which case the MMG hyperbola becomes the TMG parabola
\begin{equation}\label{halfp}
y^2 =4x\, . 
\end{equation}
The inequalities (\ref{ineq4}) simplify to (i) linear inequalities  in the $(x,y)$-plane, 
\begin{equation}
x<\sigma^2\, , \qquad 0<y<2\sigma\, , 
\end{equation}
which exclude the entire plane unless $\sigma>0$, plus (ii) the (no-ghost) restriction $\sigma<0$ on the parabola parameters. These are contradictory requirements,  so {\it the unitarity conditions cannot be satisfied}. 

\subsection{Unitary MMG}

The inequalities  (\ref{ineq4}) imply that
\begin{equation}
y-\alpha x < 2\sigma < \alpha x- 2\alpha \sigma^2\, , 
\end{equation}
and hence that 
\begin{equation}
y<2\alpha \left(x-\sigma^2\right)\, . 
\end{equation}
But this contradicts the requirements that $x<\sigma^2$ and $y>0$ unless
\begin{equation}
\alpha <0\, . 
\end{equation}
Equivalently, unitarity requires $\gamma>0$, in agreement with the result of \cite{Bergshoeff:2014pca}. 

\begin{figure}
\includegraphics[width=\textwidth]{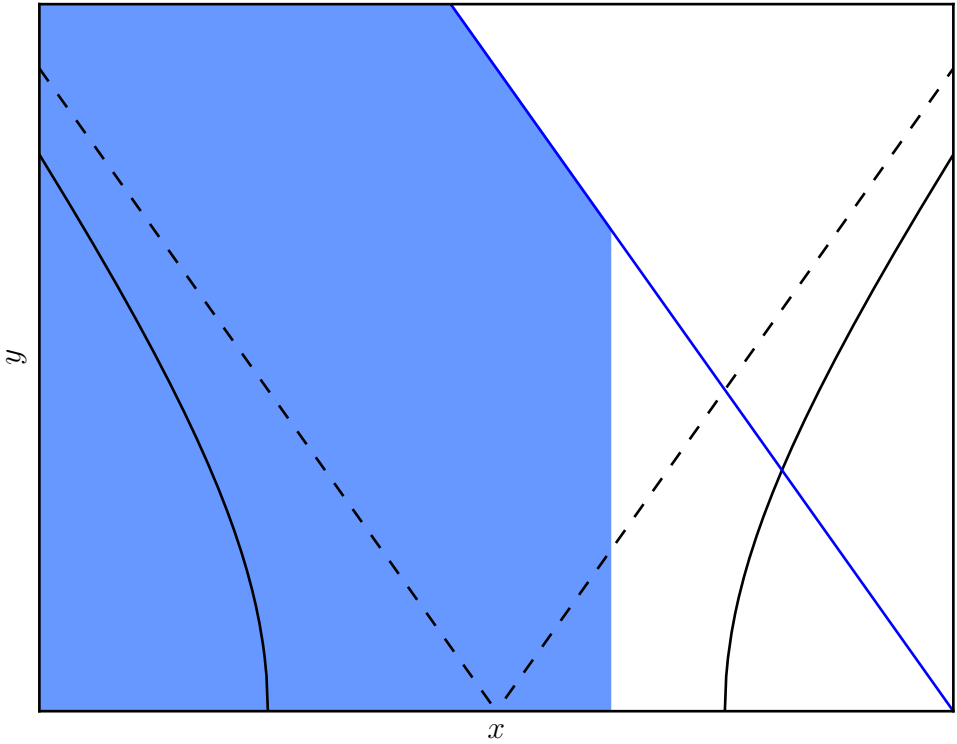}
\caption{The MMG hyperbola ($y>0$) for $(1 + \sigma \alpha)>0$ and $\sigma(2 + \sigma \alpha)>0$. The region defined by the inequalities (\ref{ineq4}) is shaded in blue. The chiral point is the intersection of the hyperbola with the solid blue line, which is parallel to the left asymptote.\label{ychpve}}
\end{figure}

\begin{figure}
\includegraphics[width=\textwidth]{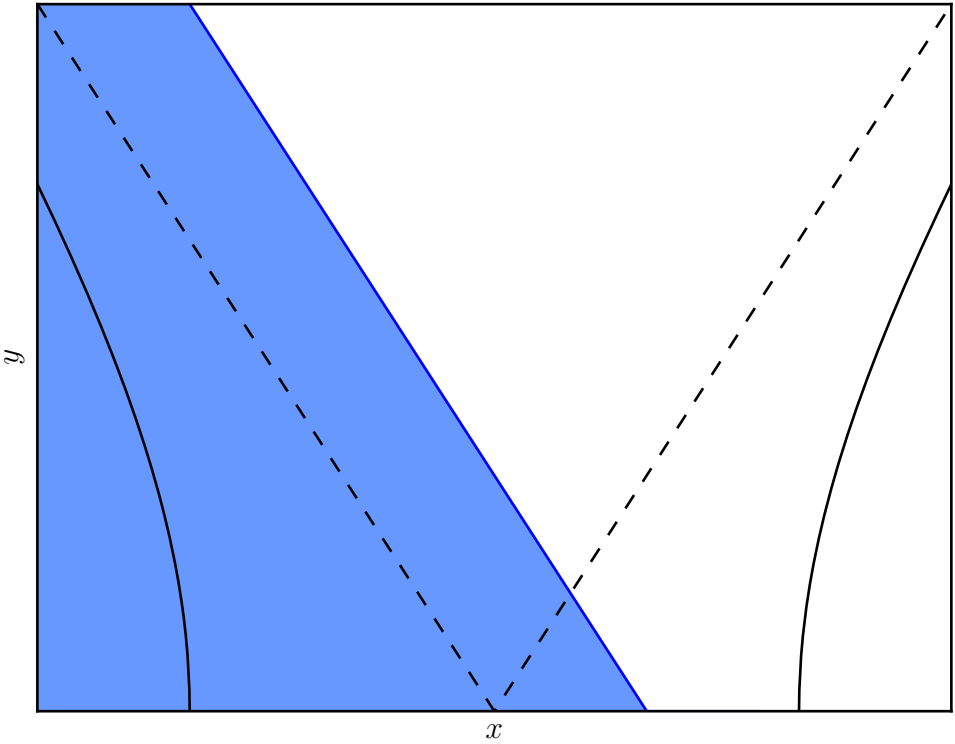}
\caption{The MMG hyperbola ($y>0$) for $(1 + \sigma \alpha)>0$ and $\sigma(2 + \sigma \alpha)<0$. The region defined by the inequalities (\ref{ineq4}) is shaded in blue. Note that the right-hand branch is now excluded outright; the chiral point is at $y_{\rm ch}<0$ on this excluded branch.\label{ychnve}}
\end{figure}

Given $\alpha<0$, the second inequality of (\ref{ineq4}) can be written as another upper bound on $x$:
\begin{equation}
x< \frac{2\sigma}{\alpha} \left(1+\sigma\alpha\right) \equiv x^*\, , 
\end{equation}
and this is a stronger upper bound than $x<\sigma^2$ when $\sigma(2+\sigma\alpha)>0$. Otherwise it is weaker, so 
\begin{itemize}

\item $x< x^*$ if $\sigma(2+\sigma\alpha)>0$. In this case $y_{\rm ch}>0$ but, since $x^*<\sigma^2$, the chiral point is outside the allowed region. See Figure \ref{ychpve}.

\item $x<\sigma^2$ if $\sigma(2+\sigma\alpha)<0$. In this case $y_{\rm ch}<0$,  so the chiral point is again outside the allowed region. See Figure \ref{ychnve}. 
\end{itemize}
We learn from this that {\it the chiral point is always outside the allowed region}.

The  MMG hyperbola crosses the $x$-axis at $x=0$ and at\footnote{When $(1+\sigma\alpha)=0$ the hyperbola degenerates into its asymptotes, but this case is excluded by definition of MMG.}
\begin{equation}
x= - \frac{4(1+\sigma\alpha)}{\alpha^2} \equiv x_{\rm int} \, . 
\end{equation}
For future use we record here that
\begin{equation}\label{xrelation}
x_{\rm int}-x^*= \frac{x_{\rm int}}{2} (2 + \sigma\alpha) \,, \qquad x_{\rm int} -\sigma^2= - \frac{(2+\sigma\alpha)^2}{\alpha^2} <0\, . 
\end{equation}

The above facts suggest that we should organize the analysis according to the signs of  $(1+\sigma\alpha)$ and $\sigma (2 + \sigma \alpha)$. As observed earlier, we have seen that we may choose either sign for 
$(1+\sigma\alpha)$ without loss of generality but we will ignore this fact, for now,  in order  to simplify  comparison with the results of \cite{Bergshoeff:2014pca}. This means that we have four generic cases to consider, plus the two special cases for which $\sigma = 0$ or $\sigma \alpha = -2$. Each of the four generic cases corresponds (given $\alpha <0$) to a definite sign of $\sigma (1 +\sigma \alpha)$ which we will also need to know.

\begin{enumerate}
\item $(1+\sigma\alpha)>0 \ {\rm and} \,  \sigma(2+\sigma\alpha)>0$. In this case $x_{\rm int}<0$, so the right-hand branch of the hyperbola is the one passing through the origin. The chiral point is on this branch. The two conditions imply
\begin{equation}
|\alpha|^{-1} > \sigma>0 \,  \qquad \left(\Rightarrow \ \sigma(1+\sigma\alpha)>0\right). 
\end{equation}
This corresponds to the third of the possibilities deduced in \cite{Bergshoeff:2014pca}.

The allowed region in the $(x,y)$-plane lies to the left of the line $x=x^*$, which excludes the entire  right-hand branch of the hyperbola. But $x_{\rm int}<x^*$, so the entire upper-half of the left-hand branch is inside the allowed region. 

\item $(1+\sigma\alpha)>0\ {\rm and}\, \sigma(2+\sigma\alpha)<0$. Again $x_{\rm int}<0$, so the right-hand branch passes through the origin. This case occurs when
\begin{equation}
\sigma<0 \, , \qquad \left(\Rightarrow \ \sigma(1+\sigma\alpha)<0\right). 
\end{equation}
This corresponds to the first of the possibilities deduced in \cite{Bergshoeff:2014pca}. 

The allowed region now  lies to the left of the line $x=\sigma^2$, and under the line $y-\alpha x=2\sigma$, 
which now passes through the chiral point  on the {\it lower-half} of the right-hand branch of the hyperbola. This again means that the entire right-hand branch is
excluded while the entire upper-half of the left hand branch is allowed.  

\item $(1+\sigma\alpha)<0\ {\rm and}\, \sigma(2+\sigma\alpha)<0$. In this case $x_{\rm int}>0$, so the left-hand branch of the hyperbola is now the one passing through the origin. This case is only realised when
\begin{equation}
\sigma> 2|\alpha|^{-1}\, , \qquad \left(\Rightarrow \ \sigma(1+\sigma\alpha)<0\right). 
\end{equation}
This corresponds to part of the second possibility deduced in \cite{Bergshoeff:2014pca}.

The allowed region lies to the left of the line $x=\sigma^2$, but since $x_{\rm int} <\sigma^2$, the entire upper-half of the left-hand branch of the hyperbola is in the allowed region. The entire right-hand branch is excluded because  the line $y-\alpha x=2\sigma$  passes through the chiral point  on the {\it lower-half} of the right-hand branch of the hyperbola. 

\item $(1+\sigma\alpha)<0\ {\rm and}\, \sigma(2+\sigma\alpha)>0$. Again $x_{\rm int}>0$ so the left-hand branch of the hyperbola passes through the origin. This case occurs when
\begin{equation}
2|\alpha|^{-1} > \sigma >|\alpha|^{-1}\, , \qquad \left(\Rightarrow \ \sigma(1+\sigma\alpha)<0\right). 
\end{equation}
This corresponds to another part of the second possibility deduced in \cite{Bergshoeff:2014pca}.

The allowed region lies to the left of the line $x=x^*$. But $x_{\rm int}>x^*>0$, so the entire right-hand branch of the hyperbola is excluded while the left-hand branch is allowed. \end{enumerate}
Notice that $\sigma(2+\sigma\alpha)$ is $H$-inert, so cases $1\&4$ and cases $2\&3$ are both ``$H$-pairs''.  In all cases only the upper half of the left-hand branch is allowed by unitarity. 

To summarise, cases 1\&2 cover $(1 + \sigma \alpha)>0$ {\it except} $\sigma =0$ and cases 3\&4 cover $(1 + \sigma \alpha)<0$ {\it except} $\sigma \alpha = -2$. We now deal with the two special cases $\sigma =0$ and $\sigma \alpha = -2$. As these are mapped into each other by  $H$-duality, we can limit the discussion to $\sigma =0$ without loss of generality. At $\sigma=0$ the previous discussion goes through except that the two figures degenerate into a single figure with the chiral point moving to the origin and the ``chiral boundary line'' becoming $y=\alpha x$.  

The upshot is that the four cases have now been reduced to just {\it two} cases, related by $H$-duality. These are
\begin{equation}
\sigma < \frac{1}{|\alpha|} \quad \& \quad \sigma > \frac{1}{|\alpha|} \, , 
\end{equation}
and in both cases only the upper half of the left-hand branch of the MMG hyperbola  is allowed by unitarity. 

\subsection{The left-hand branch restriction}

The results of \cite{Bergshoeff:2014pca} for the regions in the MMG parameter space allowed by unitarity have been recovered, and extended to include the $\sigma =0$ case. However, each of the three allowed parameter regions  found in \cite{Bergshoeff:2014pca} came with a sign restriction. What we now aim to show is that this sign restriction is equivalent to the statement that the allowed segment of the MMG hyperbola is on its left-hand branch. 

A point on the MMG hyperbola is on the left hand branch if and only if
\begin{equation}
x< \left\{\begin{array}{ccc}  x_{\rm int}& {\rm if}& (1+\sigma\alpha)>0 \\ 0&{\rm if} & (1+\sigma\alpha)<0\, , \end{array}\right.
\end{equation}
but a sufficient conditon that applies in either case is $x<x_{\rm int}/2$, or
\begin{equation}
\alpha^2 x + 2(1+\sigma\alpha) <0\, . 
\end{equation}
Using the definition of $x$ given in (\ref{xandy}), we may rewrite this inequality as 
\begin{equation}\label{signineq}
\Lambda+ \alpha\Lambda_0 - \frac{2(1+\sigma\alpha)^3}{\alpha^2}\, \mu^2 >0\, . 
\end{equation}
On the other hand, equation (\ref{quad}) viewed as a quadratic equation for $\Lambda+\alpha\Lambda_0$ has the solution\footnote{This agrees with the corresponding equation of 
\cite{Bergshoeff:2014pca} if one sets $\sigma^2=1$, as was done there.}
\begin{equation}
(\Lambda+ \alpha\Lambda_0)/\mu^2 = -\frac{2\sigma(1+\sigma\alpha)^3}{\alpha} \left[1 \mp \sqrt{1+ \frac{\alpha\Lambda_0/\mu^2}{\sigma^2(1+\sigma\alpha)^2}}\right]\, . 
\end{equation}
The sign choice appearing here enters into the unitarity analysis of \cite{Bergshoeff:2014pca}. To see the implications of this sign, we substitute for $\Lambda+\alpha\Lambda_0$ in 
(\ref{signineq}) to deduce the inequality 
\begin{equation}
\mp \sigma(1+\sigma\alpha) \sqrt{1+ \frac{\alpha\Lambda_0/\mu^2}{\sigma^2(1+\sigma\alpha)^2}} \ > -\frac{(1 + \sigma \alpha)^2}{\alpha} >0 \,.
\end{equation}
As the left hand side must be positive, we conclude that the sign choice is correlated with the sign of $\sigma(1+\sigma\alpha)$:

\begin{itemize}
\item $\sigma(1+\sigma\alpha)>0$ requires the {\it bottom sign}. This applies to case 1.

\item $\sigma(1+\sigma\alpha)<0$ requires the {\it top sign}. This applies to cases 2, 3 and 4. 
\end{itemize}
This result agrees with \cite{Bergshoeff:2014pca}. Cases 3\&4, together with the special case $\sigma \alpha = -2$,  jointly correspond to the second possibility deduced in \cite{Bergshoeff:2014pca}. Cases 1\&2 correspond to the other two possibilities found in \cite{Bergshoeff:2014pca}.

Because of the $H$-duality, we lose no generality by focusing on the cases $1\&2$ if we  also allow  for $\sigma=0$. Then $(1+\sigma\alpha)>0$ and
every point on the left-hand MMG hyperbola with $y>0$ has $x< x_{\rm int}$. Proceeding as before we now find that
\begin{equation}\label{proceed}
\mp \sigma(1+\sigma\alpha) \sqrt{1+ \frac{\alpha\Lambda_0/\mu^2}{\sigma^2(1+\sigma\alpha)^2}} \ > -\frac{(1+\sigma\alpha)(2+\sigma\alpha)}{\alpha}\, , 
\end{equation}
and by squaring we deduce that 
\begin{equation}\label{lamzerorange}
\Lambda_0/\mu^2 <  -\frac{4(1+\sigma\alpha)^3}{|\alpha|^3}\, . 
\end{equation}
Each choice of $\Lambda_0$ satisfying this inequality corresponds to one point on the upper left-hand branch of the MMG hyperbola.  We also learn that 
\begin{equation}
1+ \frac{\alpha\Lambda_0/\mu^2}{\sigma^2(1+\sigma\alpha)^2} > \frac{(2+\sigma\alpha)^2}{\sigma^2\alpha^2} >0 \, , 
\end{equation}
which confirms the reality of the square root in the expression on the left hand side of (\ref{proceed}).

\section{The flat-space limit}

Recall that $y= 2/(\mu\ell)$.  This shows that the points at which the MMG hyperbola crosses the $x$-axis correspond to flat space limits of AdS$_3$. There are two such points, one on the right-hand branch and another on the left-hand branch. The flat space limit on the left-hand branch is on the boundary of the region allowed by bulk and boundary unitarity. Strictly speaking, this boundary point\footnote{On the hyperbola; it is actually a surface
parametrised by $(\sigma,\alpha)$ in the full parameter space; this becomes a curve after a choice of normalisation of parameters.} is not in the allowed region. Although the bulk unitarity conditions have a smooth flat-space limit, the asymptotic symmetry algebra changes.

It has been argued in \cite{Bagchi:2012cy} that the holographic dual to a 3D bulk gravity theory with flat-space asymptotics should be a theory invariant under the 2D ``Galilean Conformal Algebra'' (GCA). This is a Wigner-In\"{o}n\"{u} contraction of the more usual Virasoro$\,\oplus\,$Virasoro symmetry algebra. The {\it non-zero} commutators of the GCA 
are
\begin{eqnarray}\label{GCA}
\left[L_m,L_n\right] &=& (m-n) L_{m+n} + \frac{c_1}{12} \left(m^3-m\right) \delta_{m+n} \nonumber \\
\left[ M_n , L_n \right]  &=& (m-n) M_{m+n} + \frac{c_2}{12}\left(m^3-m\right) \delta_{m+n}\, , 
\end{eqnarray}
and the central charges $c_1$ and $c_2$ are limits of linear combinations of the central charges $c_\pm$ of the 2D conformal algebra:
\begin{equation}
c_1 = \lim_{\ell\to\infty}\left(c_+-c_-\right) \, , \qquad c_2 = \lim_{\ell\to\infty}\left[ \ell^{-1}\left(c_+ +c_-\right)\right] \, . 
\end{equation}
 
For MMG we have, using the formula (\ref{cpm1})  and the fact that $y\to 0$ as $\ell\to\infty$ for fixed $\mu$, 
\begin{equation}
c_1= \frac{3}{\mu G_3}\, , \qquad c_2 = \frac{3}{G_3} \left(\sigma + \frac{\alpha \bar x}{2}\right)\, , 
\end{equation}
where $\bar x$ is one of the two values of $x$ on the MMG hyperbola when $y=0$; i.e. {\it either} $\bar x=0$ {\it or}  $\bar x= x_{\rm int}$. 
As explained in subsection \ref{subsec:duality}, we may choose the MMG parameters so that $(1+\sigma\alpha)>0$ without loss of generality. Making this choice 
(because the TMG limit is then simple) we have $x_{\rm int}<0$,   so that $\bar x=0$ on the right-hand branch of MMG hyperbola and $\bar x= x_{\rm int}$ on 
the left-hand branch.  Thus
\begin{equation}
c_2=3\sigma/G_3  \qquad ({\rm right-hand\ branch})\, . 
\end{equation}
As only the right-hand branch survives (as a parabola) in the TMG limit, we deduce that $c_2=3\sigma/G_3$ for flat-space TMG, in agreement with \cite{Bagchi:2012yk}. 
On the left-hand branch we have 
\begin{equation}
c_2 = -\frac{3}{\alpha G_3}(2+ \sigma\alpha) \qquad ({\rm left-hand\ branch})\, . 
\end{equation}
Not only is this a new possibility, not available to TMG, but also it is the value of $c_2$ found by choosing the flat-space limit on the boundary of the segment of the MMG hyperbola allowed by unitarity. 

It was argued in  \cite{Bagchi:2012cy} that, under certain conditions, the 2D GCA algebra admits a unitary truncation to one copy of the Virasoro algebra and the boundary theory becomes the chiral half of a CFT. Specifically, the conditions are that  
\begin{equation}\label{conditions}
c_1>0\, , \qquad c_2=0\, , \qquad \Delta=0\, , 
\end{equation}
where $\Delta$ is the eigenvalue of the $M_0$ generator, equal to the graviton mass. In \cite{Bagchi:2012yk} these conditions were satisfied by taking a $\sigma\to0$ limit in which flat-space TMG degenerates to 3D conformal gravity. As just observed, MMG has {\it two} flat-space limits, one on the left-hand branch of the MMG hyperbola and one on the right-hand branch, but the condition $c_2=0$ cannot be satisfied on the left-hand branch. However, setting $\sigma =0$ makes both $c_2$ and $\Delta$ vanish at the flat-space limit on the right-hand branch. At this point, the MMG equation degenerates to
\begin{equation}\label{flat}
C_{\mu\nu} + \frac{\gamma}{\mu}J_{\mu\nu}=0\, . 
\end{equation}
Taking the trace we deduce that 
\begin{equation}
R^{\mu\nu}R_{\mu\nu} -\frac{3}{8}R^2 =0\, . 
\end{equation}
The left hand side of this equation is the curvature-squared scalar of New Massive Gravity (NMG) \cite{Bergshoeff:2009hq}. There are some similarities here with  ``New Topologically Massive Gravity'' (NTMG)  \cite{Andringa:2009yc,Dalmazi:2009pm} for which the field equation is the same as (\ref{flat}) but with the $K$-tensor of NMG replacing the $J$-tensor of MMG; as observed in \cite{Arvanitakis:2014yja}, both tensors have the same scalar trace. The $J$-tensor does not contribute to linear order in the equation (\ref{flat}), so its linearisation (about flat space) is just linearised 3D conformal gravity, which does not propagate
any modes. In contrast, the non-linear theory does have local degrees of freedom; a slight modification of the Hamiltonian analysis of \cite{Bergshoeff:2014pca} (to allow for $\sigma=0$) shows that the physical phase-space of the non-linear theory has dimension $2$ per space point. There is therefore a linearisation instability of the MMG model with field equation (\ref{flat}),   analogous to the linearisation instability of NTMG \cite{Hohm:2012vh}.  

Linearisation instabilities of this type occur when the linearised theory has a gauge invariance that is explicitly broken by interactions in the full theory. This  phenomenon does not occur for TMG (the linearisation instability at the chiral point  discussed in \cite{Maloney:2009ck} is of another type) but it does occur
for NMG at partially massless AdS vacua \cite{Blagojevic:2011qc}, where the accidental gauge invariance of the linearised theory is again linearized Weyl invariance.  The  (unique) AdS vacuum at the ``merger point''  of MMG is ``partially massless'', implying a linearisation instability in that case too \cite{Arvanitakis:2015yya}; in fact, equation 
(\ref{flat}) is the flat space limit of the MMG equation at its merger point.

\section{Discussion}

We have presented a geometrical re-analysis of the bulk and boundary unitarity condition found in \cite{Bergshoeff:2014pca} for the 3D ``minimal massive gravity'' (MMG) model with AdS$_3$ asymptotics. The four (un-normalized) dimensionless parameters of the MMG action are considered in two pairs; one pair (essentially, the cosmological constant and cosmological parameter in units of the square of the TMG mass parameter) parametrises a plane. The unitarity conditions are linear inequalities in the plane but an AdS$_3$ vacuum exists only for points lying on a hyperbola  in this plane. We have called this the MMG hyperbola; it degenerates to a parabola in the TMG limit.  Both the linear inequalities and the hyperbola  depend on the other two parameters, which are the coefficient $\sigma$ of the Einstein-Hilbert term and the coefficient $\alpha$ of the new MMG term, which are restricted by the condition $(1+\sigma\alpha)\ne0$.  The linear inequalities are inconsistent unless $\alpha<0$, in which case they 
determine an ``allowed'' region in the plane. In all cases, only the upper left-hand branch of the  MMG hyperbola is in this region, and this corresponds to a sign choice  in the quadratic  equation that gives the AdS radius in terms of a cosmological parameter $\Lambda_0$, which is restricted only by a simple inequality. 

In our analysis, the values of $\sigma$ for which $\sigma(2+\sigma\alpha)=0$ are special, and this leads to four cases, depending on the signs of $\sigma(2+\sigma\alpha)$ and $(1+\sigma\alpha)$.  Only three cases arose in the analysis of  \cite{Bergshoeff:2014pca} because $\sigma=0$ was excluded but $\sigma\alpha=-2$ was not singled out as special. Once both the $\sigma=0$ and  $\sigma\alpha=-2$ cases are included, the number of connected parameter regions allowed by unitarity is reduced to two  which differ according to the sign of $(1+\sigma\alpha)$.  However, we have shown that any MMG model with $(1+\sigma\alpha)>0$  is equivalent to some MMG model  with $(1+\sigma\alpha)<0$, and {\it vice versa}. We may therefore choose either sign for  $(1+\sigma\alpha)$ without loss of generality. The choice $(1+\sigma\alpha)>0$ has the advantage that it simplifies the TMG limit, and the net result for this choice  is that the only restrictions on the parameter space imposed by unitarity (given AdS$_3$ asymptotics) 
are, up to equivalence, 
\begin{equation}
\alpha <0\,, \qquad \sigma< \frac{1}{|\alpha|}\, ,  \qquad \Lambda_0/\mu^2 < \frac{4(1+\sigma\alpha)^3}{\alpha^3}\, . 
\end{equation}
This defines {\it a single connected region in the space of parameters of the MMG action}\footnote{It corresponds to some  connected subspace of the parameters $(\bar\sigma,\gamma,\bar\Lambda_0)$ of the MMG equation; $\alpha<0$ implies $\gamma>0$  but to find the analogous explicit restrictions on $\bar\sigma$ and $\bar\Lambda_0$ would require an explicit inversion of the map defined by equations (\ref{bars})  and (\ref{barlamzero}), and this is known only  for $\Lambda_0=0$ \cite{Arvanitakis:2014yja}.}. It excludes  TMG. 

If we fix the scaling invariance (induced by a rescaling of the 3D Newton constant) by choosing $\alpha=-1$ then the above unitarity restriction reduces (up to equivalence) to 
\begin{equation}
\sigma<1\, ,   \qquad \Lambda_0/\mu^2 < -4(1-\sigma)^3 \, , \qquad \left(\alpha=-1\right)\, . 
\end{equation}
Of course, this is only a necessary condition for the  unitarity of any 2D CFT dual to MMG, but that is all we can test at the level of  semi-classical effective 3D gravity, and it is a test that TMG fails but MMG passes.

We have also used our results  to investigate Minkowski vacua in the context of the particular proposal for flat-space holography put forward in \cite{Bagchi:2012cy}, and to compare with the situation for TMG. 
One new feature of MMG is the existence of two flat space limits. One is a Minkowski limit of an MMG model with AdS asymptotics satisfying all unitarity conditions, but this does not appear to be useful. Instead, the flat-space limit that is available to both TMG and MMG is the one for which proposed unitarity conditions on central charges of the GCA algebra can be satisfied. In this limit TMG degenerates to 3D conformal gravity but MMG degenerates to a very different theory with local degrees of freedom, although this are not apparent from the linearised theory. It thus appears that MMG has implications not only for AdS holography but also for flat-space holography.  

\section*{Acknowledgements}
A.S.A. and P.K.T. acknowledge support from the UK Science and Technology Facilities Council (grant ST/L000385/1). A.S.A. also acknowledges support from Clare Hall, Cambridge,  and from the Cambridge Trust. We are grateful to Eduardo Casali, Elias Kiritsis and Alasdair Routh for helpful discussions, and to Eric Bergshoeff and Wout Merbis for helpful correspondence. 

\bigskip


\providecommand{\href}[2]{#2}\begingroup\raggedright\endgroup

\end{document}